\documentclass[letterpaper]{mn2e}
\begin{document}

\input epsf

\title[E-E Mergers and the thickening of the FP]{Mergers between elliptical galaxies \\ and the thickening of the fundamental plane}
\author[A.C. Gonz\'alez-Garc\'{\i}a \& T.S. van Albada]{A.C. Gonz\'alez-Garc\'{\i}a$^1$ and T.S. van Albada$^1$ \\
$^1$ Kapteyn Astronomical Institute,P.O. BOX 800, 9700 AV Groningen, The Netherlands}

\maketitle

\begin{abstract}
We have carried out computer simulations to study the effect of merging on the fundamental plane (FP) relation. Initially, systems are spherical Jaffe models following a simple scaling relation ($M/R_e^2 = constant$). They have been put on the FP by imposing different {\it M/L} values. Various orbital characteristics have been considered. Our results show that the merger remnants lie very close to the FP of the progenitors. Although non-homology is introduced by the merging process, mergers among homologous galaxies leave a pre-existing FP-relation intact. As a side result we find that variations in the point of view lead to non-negligible scatter about the FP.
\end{abstract}

\section{Introduction}

The radii, velocity dispersions and surface brightnesses of elliptical galaxies follow a relation known as the Fundamental Plane (hereafter FP):

\begin{equation}
	\log R_e = \alpha \log \sigma_o + \beta \; SB_e + \gamma ,
\label{eq:1}
\end{equation}
where $R_e$ is the effective radius, $\sigma_o$ is the central velocity dispersion and $SB_e = -2.5 \log \langle I \rangle_e $, with $\langle I \rangle_e  = L/2\pi R_e^2$ (Dressler et al. 1987; Djorgovski and Davis 1987; Bender, Burstein and Faber 1992). 

The existence of the FP-relation (eq. \ref{eq:1}) is, at least in principle, easy to understand: it follows from the condition of dynamical equilibrium (`virial theorem'), and the regular behaviour of both mass-to-light ratio and structure (approximate homology) along the luminosity sequence (Faber et al. 1987; Gunn 1989; Renzini and Ciotti 1993). The detailed implications of the observed values of the coefficients in eq. \ref{eq:1} remain somewhat of a mystery however.

The question we will address below is the remarkable tightness of the FP-relation in the light of galaxy encounters and mergers. Although the frequency of galaxy encounters in our neighbourhood is low, encounters and mergers appear to have been frequent in the recent past. This is illustrated for example by the large number of merging ellipticals in a cluster at $z=0.83$ (van Dokkum et al. 1999). These authors conclude that about $50\%$ of present day cluster ellipticals experienced a major merger at $z<1$.

One may ask, therefore, what effect encounters and mergers will have on the FP-relation, eq. \ref{eq:1}. If pre-merger galaxies follow the FP, will the end-product of the mergers also lie on the FP? In other words, is the FP `conserved' by encounters and mergers? Or, alternatively, is the FP, in part perhaps, the outcome of frequent encounters and mergers?

We have studied these questions with N-body simulations of tidal encounters and mergers. A survey has been carried out for a large range of galaxy and orbital parameters. This survey and its results will be addressed in a forecoming paper. Here we use the information relevant for the FP.

Our approach is as follows. We simulate encounters for a range of galaxy and orbital parameters. The galaxy parameters are scaled in such a way that prior to the encounter galaxies follow a FP-relation. We then measure the relevant parameters after the tidal encounter or merger and investigate how the pre-existing FP-relation is altered.

Recent work on galaxy mergers has focused on spiral galaxies (for a recent review see Naab \& Burkert 2001), and work on mergers of elliptical galaxies has usually been done for purposes other than the observed FP. Levine \& Aguilar (1996) performed semi-analytic calculations to see how encounters in a cluster may affect the FP.

A number of recent papers deal with the fate of satellite galaxies with dense cores, when they sink into larger elliptical galaxies, and whether or not this preserves the so called `core-FP' (Weinberg 1997; Holley-Bockelmann \& Richstone 1999, 2000; Merritt \& Cruz 2001; Milosavljevic \& Merritt 2001). The core-FP makes use of the same observables as the FP but is based on a smaller aperture. If a dense satellite preserves its core when sinking into the inner parts of the elliptical galaxy this may alter the position of the galaxy on the core-FP. To preserve the core-FP, the dense satellite core must somehow be destroyed. It has been argued that black holes may play a role in this. The regular FP will also be affected, but is not as sensitive to sinking satellites as the core FP.

A paper with a similar aim as ours is that of Capelato, de Carvalho and Carlberg (1995, hereafter CCC). CCC argue that the FP can be understood as a sequence of non-homologous systems with essentially constant $M/L$. However, their progenitors have soft cores (King models with $\log r_t/r_c=1.03$),  and deviate substantially from real ellipticals. Moreover, violent relaxation during the interaction may alter the initial density profile, although such an effect is not observed by CCC. This has led us to study the effect of mergers on the FP again, with progenitors following the $R^{1/4}$ law.

\section{The FP-relation and the virial theorem}

For galaxies in equilibrium a relation analogous to equation \ref{eq:1} can be derived from the virial theorem (hereafter VT). This relation will serve as the basis for the comparison of our model galaxies with the observed FP.

Let $E_{\rm T}$ and $E_{\rm W}$ be the kinetic and potential energies of a (model) galaxy with mass $M$. Define the velocity dispersion $\langle v^2 \rangle$ and the gravitational radius $R_{\rm G}$ through $M\langle v^2 \rangle=2E_{\rm T}$ and $GM^2/R_{\rm G}=-E_{\rm W}$. Then, according to the VT: $\langle	v^2 \rangle=GM/R_{\rm G}$.
Following the notation of CCC we introduce the structural constants $C_v$ and $C_r$ such that

\begin{equation}
	C_v \equiv \langle v^2 \rangle/\sigma_o^2 ,
\end{equation}
\begin{equation}
	C_r \equiv R_{\rm G}/R_e ,
\end{equation}
with $\sigma_o^2$ the projected velocity dispersion inside a fixed radius and $R_e$ the effective radius of the distribution of light. Then,

\begin{equation}
	\sigma_o^2 = GM / C_r C_v R_e.
\label{eq:5}
\end{equation}
Further, let $\langle I \rangle_e \equiv L/2\pi R_e^2 $,
with L the total luminosity, and
$ \langle \Sigma \rangle_e \equiv M / 2 \pi R_e^2$.
Then: $\langle \Sigma \rangle_e / \langle I \rangle_e = M/L $.
Equation \ref{eq:5}  can now be written in the following form:

\begin{equation}
	R_e = \frac{C_r C_v}{2 \pi G}\frac{\sigma_o^2}{\langle I \rangle _e(M/L)} ,
\label{eq:9}
\end{equation}
or,

\begin{equation}
	\log R_e = 2 \log  \sigma_o + 0.4 SB_e + \log (C_r C_v) - \log (M/L) + \gamma,
\label{eq:10}
\end{equation}
where $\gamma$ is a constant. In the following this relation will be called the VT-relation. The observed FP-relation is given by (e.g. J\o rgensen, Franx \& Kj\ae rgaard, 1996, hereafter JFK96):

\begin{equation}
	\log R_e = 1.24 \log \sigma_o + 0.328 SB_e + \gamma'.
\label{eq:11}
\end{equation}
Equation \ref{eq:10} shows that two ingredients may explain the so-called tilt of the FP with respect to the VT-relation: a systematic change of the structure `constant' $C_vC_r$ with $R_e$, (the so called `non-homology'), or a change of  $M/L$ with $R_e$. Most ellipticals appear to obey the $R^{1/4}$ law. Therefore, non-homology is probably weak.
On the other hand, the luminosities of elliptical galaxies correlate with colour as well as metallicity. This indicates that $M/L$ probably varies with {\it L}. For discussions of these issues see Renzini \& Ciotti, (1993).

\section{N-body simulations}
\subsection{The sample}

We have done a survey with the intention to map the input parameters of our systems before the interaction or merger on to the parameters of the final systems. In this survey we have several models leading to mergers (those used in this chapter) and others involving different degrees of interaction.
To build our initial models we used Jaffe (1983) models (which are spherical and isotropic) as initial conditions, using an algorithm developed by Smulders \& Balcells (1995). 

Initial models consist of at least $10^4$ particles. Although the number of particles is rather low, it is sufficient for the observables of interest. Model parameters are given in Tab. \ref{tab:fp1}, which lists the model name, initial mass ratio of the progenitors, initial impact parameter, D, and orbital energy of the system. Throughout the calculations we used dimensionless units such that the constant of gravity $G=1$ and mass and Jaffe radius of galaxy $P_1$ in in Tab. \ref{tabi} are equal to 1. A consistency check has been done with $10^5$ particles. This simulation has the same initial conditions as model $1e$. The results are essentially identical to the model with fewer particles. In figure 3 it lies on top of model $1e$. 

For models involving masses different than one, the progenitor was built homologously from the progenitor of mass one, following the scaling relation $M/R^2 = constant$ (`Fish's law', Fish, 1964)

To define the initial orbit we must choose values for the impact parameter, the orbital energy  and the initial separation. Three values for the impact parameter were used: zero (head-on collision), half of the radius enclosing the `total mass' of the larger galaxy, and this `total mass' radius (empirically determined as the radius enclosing $99\%$ of the mass of the galaxy). Both parabolic and hyperbolic encounters were considered. Elliptical orbits always lead to merger (Binney \& Tremaine $1987$, fig. $7-9$) and were not included. The distance between the progenitors at the beginning of the run was set equal to the sum of the radii of both systems plus the diameter of the smaller system (this is: $R_1+R_2+2R_1$, where subscripts 1 and 2 refer to the smaller and larger systems respectively).

The simulations were run with Hernquist's version of the {\small TREECODE} (Hernquist $1990$) on a Sun Sparc station. Softening was set to one fifth of the half mass-radius of the smaller system, while the tolerance parameter was set to $0.8$. Quadrupole terms were included in the force calculation. In all the simulations energy is conserved to within $0.5 \%$, and the systems were let to evolve after merger until they were well virialized.

The merger remnants display a wide variety of morphologies. From spherical for models with a mass ratio between progenitors of $10:1$, to highly distorted systems for more equal masses, $1:1$ or $2:1$. Head-on collisions lead to prolate non-rotating ellipticals. In these systems the point of view variation of the FP parameters is important, given the fact that a cigar shape system can be viewed as spherical or highly flattened depending on the viewing angle. These systems are therefore expected to suffer the largest variation due to point of view with respect to their location on the FP-relation. Mergers with $D \neq 0$ give oblate slowly rotating systems. In these systems, a large variation with point of view is not expected.

Anisotropy has also been studied for these models, and although the progenitors are isotropic, radial and tangential anisotropies develop during the merger process, depending on the type of orbit. This is relevant for our study since anisotropy can affect the measured central velocity dispersion, also depending on the point of view, and introduce some scatter in the FP-relation.

\subsection{The Fundamental Plane set-up}

In the following sections we will study the influence of mergers on the scatter about the FP for galaxies initially on a VT-type relation, and similarly for systems placed on a FP-like relation.

The models presented here include only luminous material. Therefore, in order to place our initial models on the Fundamental Plane (FP), we have to adopt a mass-to-light ratio for the systems. In the case that we use a fixed $M/L$ for all our models, they will lie on the VT-relation,  provided that they are built homologously fulfilling the conditions underlying equation \ref{eq:10}. In this case, for an edge-on view of the VT-relation one should use: $\log R_e$ vs. $\log (\sigma_o) + 0.2 SB_e $. In order to explore the effect of merging for galaxies lying initially on the FP one must assign varying $M/L$ values to the progenitors. If the systems on the FP behave homologously we have the approximate relation (Renzini \& Ciotti, 1993): $\log(M/L) \propto 0.20 \log(M) \propto 0.17 \log(L)$.

We have taken this approach using the coefficients of the FP given by JFK96, then $ \log L \simeq 0.78 \log M + constant$. The luminosity of individual particles follows from $l = L/N$.

For the simulated systems we have measured our observables, i.e.: effective radius, mean surface brightness inside the effective radius and the central velocity dispersion, in the following way:

First, given the number of escaping particles and their luminosity, we know the total luminosity of the system (in the initial systems we have no escapers) and therefore we look, in projection, for the radius enclosing half of the luminosity. By definition, this is the effective radius. Inside this radius we calculate the mean surface brightness. In the case of the observed FP the merger remnants will have particles with different {\it m/l} ratios and so with different luminosities. Those with higher luminosity will contribute more to the central velocity dispersion. With the same criteria as before, we take $0.2 R_e$ as the radius of the diaphragm to calculate the central velocity dispersion of the particles,  luminosity weighted like:

\begin{equation}
\sigma_o^2=\frac{N_1l_1(\sigma_o^2)_1+N_2l_2(\sigma_o^2)_2}{N_1l_1+N_2l_2},
\end{equation}
where $N$ is number of particles and the indices indicate different components.

Note that the velocity dispersion of the progenitor is independent of the luminosity of its particles, but this is not the case for the remnants.

The initial parameters of the VT and FP progenitors are given in Tab. \ref{tabi}, while values for the merger remnants are shown in Tab. \ref{tab:fp1}.
 

\begin{table}
\begin{minipage}{85mm}
\caption{Parameters for progenitor models. The columns give the model name, mass, effective radius, surface brightness and central velocity dispersion. \label{tabi}}
\begin{tabular}{cccccc}
\hline
{\bf Model} &{\bf Mass}& {\bf $R_{e}$}&{\bf $(SB_e)_{\rm {VT}}$}&{\bf $(SB_e)_{\rm {FP}}$}&{\bf $\sigma_{o}$}\\

\hline
$P_1$ &1&0.61  & 0.93&   0.93   &  0.57\\
$P_2$ &2&0.87  & 0.93&    1.11 &    0.67\\
$P_3$ &3&1.06  & 0.93&    1.22  &   0.75\\
$P_5$ &5&1.37 &0.93&   1.35  &  0.85\\
$P_7$ &7&1.62 &0.93&   1.44  &  0.92\\
$P_{10}$ & 10 &1.94 &0.93&   1.53  &  1.01\\

\hline

\end{tabular}
\end{minipage}
\end{table}



\begin{table}
\begin{minipage}{85mm}
\caption{Parameters for merger models. The first column gives the model name and the next three columns give the model input parameters, i.e., the mass ratio ($M_2/M_1$), impact parameter, and orbital energy. The last four columns give the observational parameters for each of the three principal axes of the merger remnant. $M_e$ is the projected mass inside the effective radius (excluding escaping particles).\label{tab:fp1}}
\begin{tabular}{cccccccc}

\hline

{\bf Run} &{\bf $M_2:M_1$}& {\bf $D$} & {\bf $E_{orb}$} & {\bf $R_{e}$}& {\bf $M_{e}$}& {\bf $\mu_e$}& {\bf $\sigma_{o}$}\\
\hline

$1a$ & 1:1 & 0 & 0 &    1.22    &     0.94     &   1.73   &      0.56\\
      &&&&   1.51   &      0.94    &     2.20       &  0.49\\
        &&&&  1.55      &   0.94     &    2.26      &   0.47\\
$1b$ & 1:1 & 0 & 0.0625 &	 1.34   &      0.95  &       1.94   &      0.62\\
       &&&&   1.65      &   0.95     &    2.39   &      0.49\\
       &&&&   1.73       &  0.95       &  2.49   &      0.45\\
$1c$ & 1:1 & 0 & 0.250 &	 1.94     &    0.96    &     2.73  &       0.51\\
         &&&& 2.47    &     0.96 &        3.25   &      0.40\\
         &&&& 2.52     &    0.96    &     3.30     &    0.36\\
$1e$ & 1:1 & 5 & 0 &   1.26      &   0.94     &    1.80     &    0.57\\
         &&&& 1.27    &     0.94       &  1.83    &    0.53\\
         &&&& 1.45    &     0.94       &  2.11     &    0.46\\
$1h$ & 1:1 & 10 & 0  &  1.36     &    0.96    &     1.97     &    0.57\\
        &&&&  1.32     &    0.96     &    1.89      &   0.54\\
        &&&&  1.62       &  0.96   &      2.34   &      0.46\\
$2a$ & 2:1 & 0 & 0  &   1.10      &   1.42  &       1.19   &      0.77 \\       
	 &&&& 1.33      &   1.42       &  1.60   &      0.72\\
         &&&& 1.33      &   1.42     &    1.61     &    0.65\\
$2b$ & 2:1 & 0 & 0.239  &	 1.59     &    1.42     &    2.00   &    0.63\\
         &&&& 1.88      &   1.42      &   2.36      &  0.57\\
         &&&& 1.90      &   1.42     &   2.39      &  0.56\\
$2c$ & 2:1 & 0 & 0.478 &	 1.88     &    1.41      &   2.38   &   0.54\\
         &&&& 2.17  &       1.41     &    2.69   &      0.48\\
         &&&& 2.20   &      1.41      &   2.72   &      0.53\\
$2e$ & 2:1 & 5 & 0 &	 1.22  &       1.42  &       1.41    &     0.68\\
         &&&& 1.26      &   1.42    &     1.50      &   0.60\\
         &&&& 1.37    &     1.42   &      1.67       &  0.68\\
$2h$ & 2:1 & 5 & 0 &	 1.47   &      1.43&         1.81   &      0.64\\
        &&&&  1.50  &       1.43     &    1.86   &      0.62\\
         &&&& 1.73  &       1.43   &     2.17      &   0.59	\\
$3a$ & 3:1 & 0 & 0&	 1.24    &     1.90   &      1.23  &       0.74\\
         &&&& 1.42   &      1.90      &   1.54   &      0.76\\
         &&&& 1.44    &     1.90      &   1.56     &    0.68\\
$3b$ & 3:1 & 0 & 0.387&	 1.23   &      1.77 &       1.32  &       0.79\\
         &&&& 1.35   &      1.77   &      1.51     &    0.74\\
         &&&& 1.40   &      1.77   &      1.58     &    0.68\\
$3e$ & 3:1 & 8.66 & 0&	 1.58    &     1.89  &       1.76  &      0.71\\
        &&&&  1.59      &   1.89   &      1.78   &      0.57\\
        &&&&  1.77      &   1.89    &     2.01     &    0.68\\
$5a$ & 5:1 & 0 & 0 &	 1.66    &     2.80     &    1.60      &   0.75\\
         &&&& 1.80    &     2.80    &     1.77    &     0.83\\
         &&&& 1.84    &     2.80    &     1.83    &     0.76\\
$5b$ & 5:1 & 0 & 0.761&	 1.48    &     2.73     &    1.39     &    0.79\\
          &&&&1.55   &      2.73    &     1.49  &       0.85\\
          &&&&1.56    &     2.73    &     1.51     &    0.77\\
$5e$ & 5:1 & 11.18 & 0&	 1.82      &   2.85    &     1.77     &    0.78\\
     &&&&     1.82     &    2.85   &      1.77  &       0.74\\
     &&&&     2.02     &    2.85    &     1.99    &     0.73\\
$7a$ & 7:1 & 0 & 0&	 1.64    &     3.86      &   1.31  &       0.97\\
      &&&&    1.73    &     3.86  &       1.42 &        0.93\\
      &&&&    1.75    &     3.86  &       1.46    &     0.95\\
$7e$ & 7:1 & 13.23 & 0&	 1.96    &     3.81    &     1.71   &      0.84\\
      &&&&    1.98   &      3.81 &       1.74  &       0.87\\
      &&&&    2.11   &      3.81  &       1.88  &       0.92\\
$10a$ & 10:1 & 0 & 0&	 2.22    &     5.22   &      1.76   &      0.91\\
        &&&&  2.20  &     5.22   &     1.74  &       0.94\\
        &&&&  2.22    &     5.22    &     1.76  &       0.93\\
$10e$ & 10:1 & 15.81&	0& 2.30      &   5.22     &    1.82   &      0.97\\
        &&&&  2.30     &    5.22     &   1.83      &   0.88\\
        &&&&  2.46     &    5.22      &   1.97      &   0.91\\

\hline

\end{tabular}
\end{minipage}
\end{table}

\section{Results}

\subsection{The VT-relation.}

We have considered one hundred random points of view, for each of which we have computed the various observables. Our systems, as explained before, have a wide variety of morphological features. Some are prolate while others are oblate, some have radial or tangential anisotropy while others are isotropic systems, and so on. As a result, galaxies will be scattered about the VT-relation depending on the point of view.

\begin{figure}
\begin{center}
\leavevmode
\hbox{%
\epsfxsize=4cm
\epsfbox{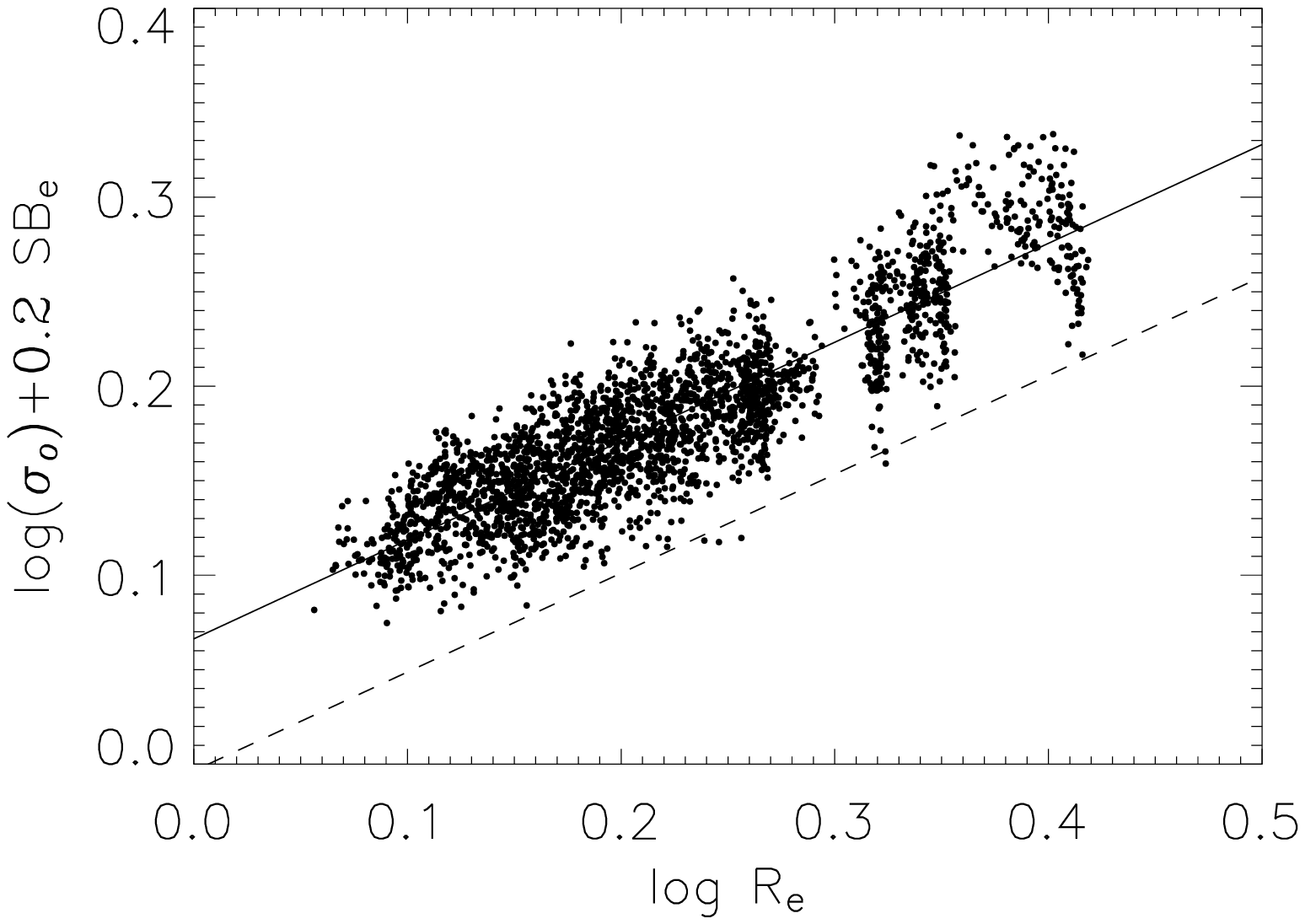}}
\hbox{%
\epsfxsize=4cm
\epsfbox{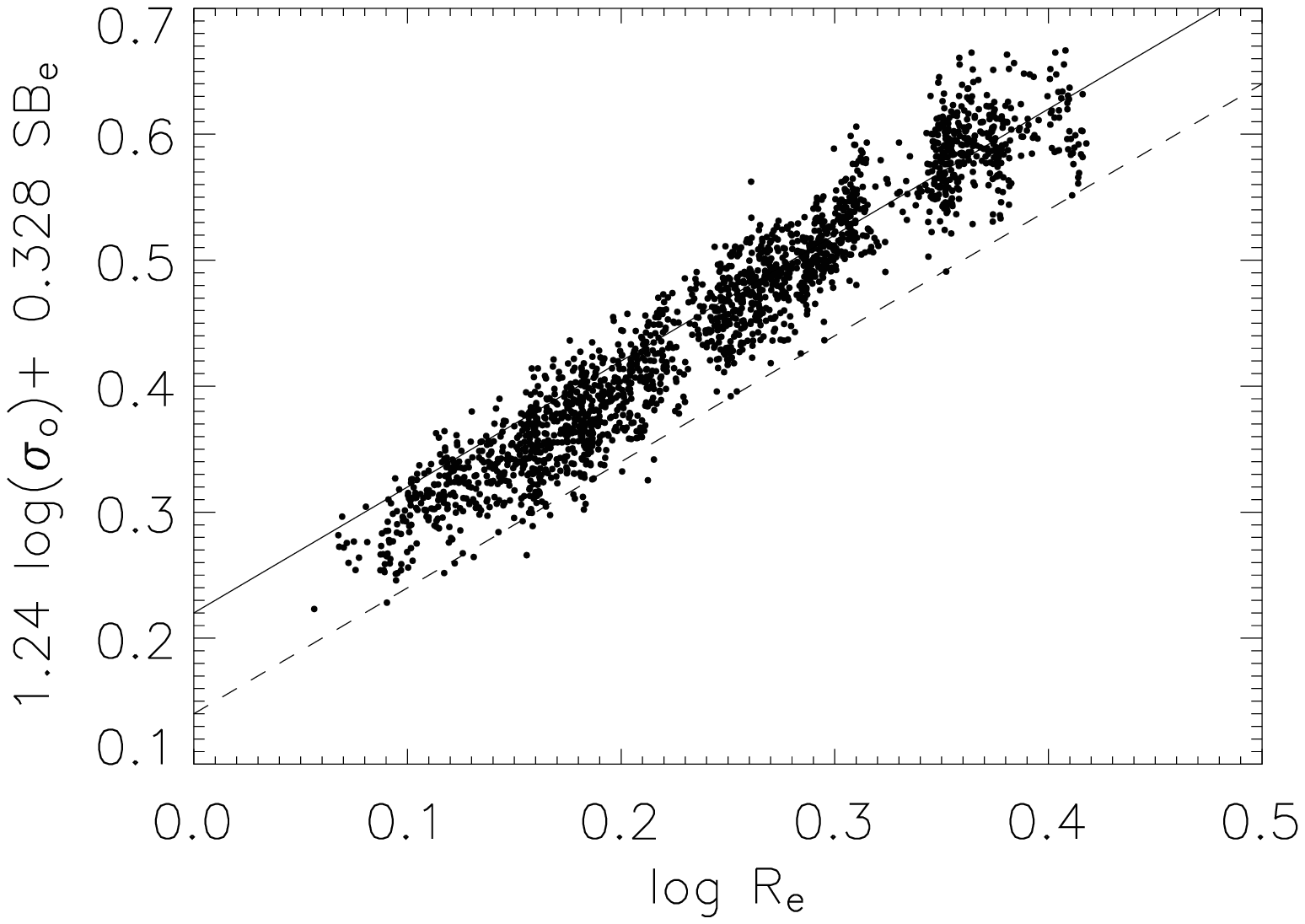}}
\caption{The effect of variations in the point of view on the scatter about the FP. Merger remnants are plotted as seen from 100 random points of view. The left panel shows our models in the VT representation. The right panel shows our models in the observed FP representation. The scatter due to projection effects is smaller than the observed scatter ( $1 \sigma$ is indicated by the dashed line), but it is not negligible. \label{fig:pv}}
\end{center}
\end{figure}

In Fig.\ref{fig:pv} we compare our results with the VT and FP relations (eq.  6 and 7). The scatter introduced by varying the point of view is not as big as the observed scatter about the FP (dashed line, using data for ellipticals in J\o rgensen et al. 1995a, 1995b), but  it is not negligible. Furthermore, because our systems do not encompass the full observed range in ellipticities, point of view effects will be larger than found here, and may well be an important source of the scatter about the FP.

\begin{figure}
\begin{center}
\leavevmode
\hbox{%
\epsfxsize=8.cm
\epsfbox{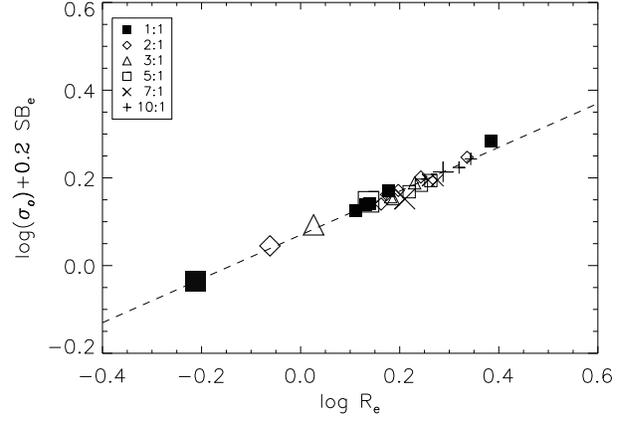}}
\caption{Location of merger remnants (small symbols) w.r.t the VT-relation. Progenitors are indicated by large symbols. The dashed line has the theoretical slope of 0.5 (see eq. \ref{eq:10}). \label{fig:fpmf}}
\end{center}
\end{figure}

Taking the mean of the cloud of points for each model we obtain Fig. \ref{fig:fpmf}. The progenitors are plotted as large symbols and the merger remnants of different mass ratios as smaller symbols. Figure \ref{fig:fpmf} shows that the merger remnants lie very close to the VT-relation of their progenitors, shown as a dashed line. This is as expected for homologous ($R^{1/4}$-law) systems with fixed $M/L$.

\begin{figure}
\begin{center}
\leavevmode
\hbox{%
\epsfxsize=8.cm
\epsfbox{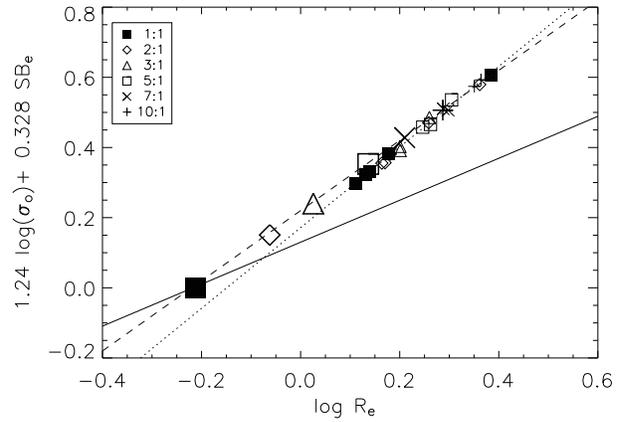}}
\caption{Location of models with respect to the observed Fundamental Plane. Large symbols indicate the initial models (i.e. the progenitors of the merger remnants). They have been put on the FP-relation of JFK96 (dashed line) by choosing appropriate values of {\it M/L}. Small symbols represent the merger remnants. They lie very close to the FP-relation but have a slightly steeper slope, indicated by the dotted line. The solid line is the VT-relation assuming constant {\it M/L}.\label{fpvd}}
\end{center}
\end{figure}

\subsection{The FP-relation.}

To examine the effect of mergers on the observed Fundamental Plane we use the FP-relation given by JFK96, eq. \ref{eq:11}.
Again, the FP parameters are calculated from 100 random points of view. Then, taking the mean for each of the systems we obtain Fig. \ref{fpvd}. The initial models are given as large symbols and the merger remnants of different mass ratios as smaller symbols. The merger remnants lie very close to the original FP but the slope is slightly steeper. Our models range from equal mass mergers to mass ratios of ten to one. This confines our sample to a small region of the FP. If we scale our models to different masses the merger remnants follow the FP, with a small scatter of the order of $0.005$ dex, smaller than the scatter introduced by projection effects ($0.024$ dex).

That the merger remnants should lie so close to the FP of the progenitors is not really surprising. It is straightforward to show that for a parabolic encounter of two equal mass $R^{1/4}$-law progenitors, the merger remnant will lie about 0.05 dex below the FP-relation in Fig. \ref{fpvd} (neglecting mass loss and assuming that the merger remnant is also an $R^{1/4}$-law galaxy).

\subsection{Homology vs. non-homology}

Our initial models, i.e. the progenitors in the merger simulations, are homologous (spherical Jaffe models) and obey Fish's (1964) law: $M_1/M_2 \propto (R_1/R_2)^2$. 
According to the formulation given in section 1 the homology constants are: $C_v \equiv \langle v^2 \rangle/\sigma^2_o $ and $C_r \equiv R_{\rm G}/R_e$. For our initial models (see also Dehnen 1993) the values for these constants are: $C_r=2.99$ and $C_v=1.70$ (due to softening the value of $C_r$ for the progenitors is somewhat smaller than the theoretical value).

\begin{figure}
\begin{center}
\leavevmode
\hbox{%
\epsfxsize=8.cm
\epsfbox{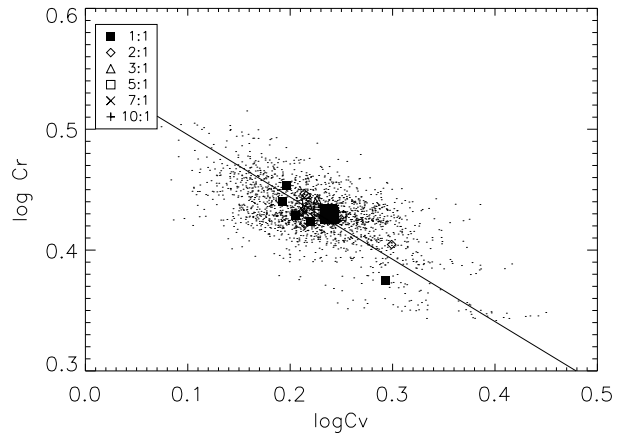}}
\caption{ Relation between structure constants $C_r$ and $C_v$. The initial models (large symbols) all lie at the same location in this plot. The merger remnants (small symbols) scatter along a line of constant $C_vC_r$. Projection effects are given by the cloud of points.\label{cvcr}}
\end{center}
\end{figure}

In Fig. \ref{cvcr} we plot the homology constants, $\log (C_v)$ vs $\log (C_r)$ . Because the progenitors are homologous they lie at the same location. Our final systems spread out in this plot, but largely along a line $\log (C_vC_r) = contant$.

Equation \ref{eq:10} shows that the approximate constancy of $C_vC_r$ minimizes deviations from the FP-relation. $C_vC_r$ is essentially a mass-ratio (total mass/projected mass), but why it should show a tendency to remain approximately constant is not clear. Mergers thus do produce some non-homology, but this non-homology is masked by the near-constancy of $C_vC_r$.

CCC carried out merger experiments of King models with concentration parameter $c = 1.03$ and fixed {\it M/L}. They find that substantial non-homology develops as a result of merging. `Second generation' mergers show a similar effect. They conclude that the FP-relation can be largely explained by the non-homologous nature of merger remnants produced by first generation mergers or even a merger hierarchy. Simulations of Nipoti, Londrillo and Ciotti (2003) also show that the FP is (marginally) conserved in a sequence of mergers.

In our simulations the progenitors are all $R^{1/4}$-law galaxies and they lie on the FP-relation as a result of varying {\it M/L}. The non-homology produced by merging is weak and the FP-relation is well conserved.

Moreover, Weinberg (2001) argues that long-term dynamical evolution 
leads to a universal profile. If this is so, merger remnants are likely to become more homologous in the course of time.

Taking these results together we conclude that the FP is insensitive to merging.

\subsection{Conclusions}

We have performed a series of simulations of mergers of spherical Jaffe models with the aim to examine what happens to elliptical galaxies on the Fundamental Plane when they merge. In other words: is the FP destroyed or conserved during the merging process?
To do so we have used purely collisionless simulations with one stellar component, and with different mass ratios for the progenitor systems. We find that mergers do not destroy the FP-relation.

As a side result from our simulations we find that projection effects lead to a small, but significant scatter, about one third of the observed scatter about the FP.

The progenitors in our simulations have the same spatial structure, but varying $M/L$ in order to put them on the observed FP. The merging process breaks the homology of the systems:
however, since $\log (C_vC_r)$ remains more or less constant, for our merger remnants this non-homology does not lead to an appreciable scatter about the FP.

\end{document}